\documentclass[fleqn,10pt]{wlscirep}
\usepackage[utf8]{inputenc}
\usepackage[T1]{fontenc}

\begin{document}

\title{Emergence of life in an inflationary universe}

\author[1,2,*]{Tomonori Totani}
\affil[1]{Department of Astronomy, School of Science, The University of Tokyo, Bunkyo-ku, Tokyo 113-0033, Japan}
\affil[2]{Research Center for the Early Universe, School of Science, The University of Tokyo, Bunkyo-ku, Tokyo 113-0033, Japan}

\affil[*]{totani@astron.s.u-tokyo.ac.jp}


\begin{abstract}
Abiotic emergence of ordered information stored in the form of RNA is
an important unresolved problem concerning the origin of life. A
polymer longer than 40--100 nucleotides is necessary to expect a
self-replicating activity, but the formation of such a long polymer
having a correct nucleotide sequence by random reactions seems
statistically unlikely. However, our universe, created by a single
inflation event, likely includes more than $10^{100}$ Sun-like
stars. If life can emerge at least once in such a large volume, it is
not in contradiction with our observations of life on Earth, even if
the expected number of abiogenesis events is negligibly small within
the observable universe that contains only $10^{22}$ stars.  Here, a
quantitative relation is derived between the minimum RNA length
$l_{\min}$ required to be the first biological polymer, and the
universe size necessary to expect the formation of such a long and
active RNA by randomly adding monomers.  It is then shown that an
active RNA can indeed be produced somewhere in an inflationary
universe, giving a solution to the abiotic polymerization problem.  On
the other hand, $l_{\min}$ must be shorter than $\sim$20 nucleotides
for the abiogenesis probability close to unity on a terrestrial
planet, but a self-replicating activity is not expected for such a
short RNA.  Therefore, if extraterrestrial organisms of a different
origin from those on Earth are discovered in the future, it would
imply an unknown mechanism at work to polymerize nucleotides much
faster than random statistical processes.
\end{abstract}


\flushbottom
\maketitle

\thispagestyle{empty}

\section*{Introduction}
In spite of recent rapid development of biology, chemistry, Earth
science and astronomy, the origin of life (abiogenesis) is still a
great mystery in science
\cite{Ruiz-Mirazo2014,Lanier2017,Szostak2017,Kitadai2018,Yamagishi2019}.
A prominent feature of life is the ordered information stored in
DNA/RNA, and how such information appeared from abiotic processes is a
crucial issue. The RNA world hypothesis
\cite{Gilbert1986,Orgel2004,Robertson2012} postulates an early era
when RNA played both the genetic and catalytic roles before the
DNA-protein world came into being. This is widely accepted due to
strong supporting evidence including catalytic activities of RNA,
especially its central role in a ribosome.  However, a more
fundamental and unsolved problem is how an RNA polymer long enough to
have a self-replicating RNA polymerase activity (i.e., RNA replicase
ribozyme) emerged from prebiotic conditions and then triggered
Darwinian evolution.

A key quantity is the minimum RNA length required to show a
self-replicating ability.  RNA molecules shorter than 25 nucleotides
(nt) do not show a specified function, but there is a reasonable
hope to find a functioning replicase ribozyme longer than 40--60 nt
\cite{Szostak1993,Robertson2012}.  RNA polymerase ribozymes produced
by {\it in vitro} experiments so far have a length longer than 100 nt
\cite{Johnston2001,Horning2016,Wachowius2019}.  Furthermore, formation
of just a single long strand may not be sufficient to initiate an
abiogenesis event.  Instead a pair of identical strands may be necessary if
one serves as a replicase ribozyme and the other as a template.

Polymerization of RNA in water is a thermodynamically uphill process,
and hence reacting monomers need to be activated.  Non-enzymatic
reactions of adding activated monomers (e.g., imidazole-activated
ribonucleotides) to an RNA oligomer have been experimentally studied
\cite{Cafferty2014,Ruiz-Mirazo2014,Kitadai2018}.  Reactions at
inorganic catalytic sites (e.g. surface of minerals such as
montmorillonite clay) may be particularly efficient
\cite{Ferris2006,JamesCleavesII2012}.  Some experiments yielded
production of up to 40-mers of RNA \cite{Huang2003,Huang2006}, which
may be long enough to have some biological activities. However, these
results have not been reproducible, and only short oligomers of up to
10-mers were produced conclusively in recent experiments, with the
abundance rapidly decreasing with the oligomer length
\cite{Kawamura1994,Joshi2009,Cafferty2014,Aldersley2017}.  This trend
is also consistent with the theoretical expectation for random adding
of monomers (see below).  An experimental difficulty is that
aggregates may easily be mistaken for polymers, depending on detection
methods \cite{Burcar2013,Cafferty2014}.

It is theoretically speculated that terminal ligation of oligomers may
hierarchically produce further longer polymers \cite{Briones2009}, but
there is no experimental or quantitative demonstration of this
starting from realistic prebiotic conditions.  A report
\cite{Costanzo2009} of experimental production of long polymers ($>
120$ nt) by ligation has been subject to reproducibility and the
aggregate/polymer discrimination problem
\cite{Burcar2013,Morasch2014,Cafferty2014}. A high concentration of
oligomers is necessary for ligation to work efficiently, but this may
be difficult because oligomer abundance rapidly decreases with
oligomer length in polymerization by monomers, even if such a ligase
activity exists.

If we consider only the conservative abiotic polymerization, i.e.,
statistically adding monomers, the probability of abiogenesis may be
extremely low on a terrestrial planet. This case is not in
contradiction with our existence on Earth, because we would find
ourselves on a planet where abiogenesis happened.  The life on Earth
is believed to have descended from the single last universal common
ancestor (LUCA) with no evidence for multiple abiogenesis events, and
we do not know any life of a different origin in the universe. The
emergence of life early in the history of Earth is often used to argue
for a high abiogenesis rate, but an arbitrarily low rate cannot be
robustly excluded \cite{Lineweaver2002,Spiegel2012,Chen2018} because
the chance probability $t_{ab}/t_\oplus \sim 0.1$ is not negligible
assuming a constant abiogenesis rate, where $t_{ab}$ is the time of
abiogenesis elapsed from the birth of Earth \cite{Pearce2018}, and
$t_\oplus$ the present age of Earth.  It may well be possible that
early Earth was a more favorable environment for abiogenesis than
present \cite{Pearce2017,Benner2019}.  There may also be an anthropic
selection effect favoring earlier abiogenesis on Earth, because an
intelligent life must emerge before the increasing solar luminosity
causes an end to Earth's habitable state (estimated to be $\sim$ 1 Gyr
in future) \cite{Lingam2019}.

For the case of a low abiogenesis rate, the number of abiogenesis
events is often considered in the Milky Way Galaxy containing about
$10^{11}$ Sun-like stars \cite{Licquia2015} or in the whole observable
universe containing $10^{22}$ stars \cite{Rudnick2003} inside a
spherical volume with a comoving radius of 46.3 Gly (or 13.8 Gly as a
light travel time distance) \cite{Spiegel2012,Sandberg2018}.  However,
the size of the observable universe is not related at all to its whole
physical size.  According to the widely accepted view of the
inflationary cosmology
\cite{Starobinsky1980,Kazanas1980,Guth1981,Sato1981,Linde1982,Albrecht1982},
the physical size of the universe created by an inflation event should
be much larger, likely including more than $10^{100}$ stars (see
below).  In that case, even if the expected number of abiogenesis
events is much less than unity in a volume size of the observable
universe, it may still be consistent with our observations provided
that abiogenesis is expected to occur somewhere in an inflationary
universe.

The aim of this work is to examine this possibility quantitatively,
assuming that the first biological RNA polymer was produced by
randomly adding monomers. Koonin \cite{Koonin2007} considered
implications of the eternal inflation theory for the origin of life.
In this scenario, most part of the universe inflates forever,
self-reproducing many subregions that undergo a conventional inflation
followed by a hot big-bang universe. Then an infinite number of stars
and galaxies would be formed, and we expect emergence of life even if
the abiogenesis probability is infinitely small. Though eternal
inflation is a theoretically likely scenario \cite{Liddle2000}, it is
difficult to confirm observationally, and a quantitative discussion is
impossible.  It is then interesting to ask if life can emerge within
the homogeneous region in which we exist, assuming its minimal size
necessary to explain observations. This work tries to give a
quantitative answer to this question.

\section*{The size of the universe created by an inflation event}

The observable universe is highly homogeneous and spatially flat on
scales many orders of magnitude larger than the causally connected
scale (horizon) in the early universe, which are called the horizon
and flatness problems, and cannot be explained by the standard big
bang cosmology. Cosmic inflation is currently the only widely accepted
solution to these problems, and furthermore, it naturally generates
scale-invariant quantum density fluctuations that serve as the seed of
galaxy formation and the large scale structure in the present
universe. Its prediction is in quantitative agreement with the
observations of the cosmic microwave background radiation anisotropy,
already constraining some theoretical models \cite{Planck2016}.

There are many models and scenarios about how inflation occurred in
the early universe \cite{Liddle2000}, but all of them consider an
epoch of exponential expansion as $a \propto \exp(H_i \, t_i)$, where
$a$ is the scale factor of the universe, $H_i$ the Hubble parameter at
the time of inflation, and $t_i$ the duration of inflation. If the
inflation occurred at the energy scale of the grand unified theory of
particle physics ($10^{16}$ GeV), $H_i$ would be about $10^{37}$
s$^{-1}$. To solve the horizon and flatness problems, the $e$-folding
number of inflation ($N_i \equiv H_i \, t_i$) must be larger than
\cite{Dodelson2003,Liddle2003} $N_{i,\min} \sim 60$.  If $N_i = N_{i,
  \min}$, a causal patch region expanded by the inflation has now the
same size as the observable universe.  It would be a fine tuning if
the inflation duration is extremely close to the minimal value to
solve the problems (i.e., $N_i - N_{i, \min} \ll N_{i,
  \min}$). Rather, we naturally expect $N_i - N_{i, \min} \gtrsim
N_{i, \min}$.  If the inflation duration is twice (three times) as
much as that required to solve the problems, the homogeneous universe
should extend $e^{60}$ ($e^{120}$) times as much as the currently
observable universe, which is $10^{78}$ ($10^{156}$) times as large in
volume, thus including about $10^{100}$ ($10^{178}$) stars.

\section*{Poissonian RNA polymerization}

Here we consider a cycle of RNA polymerization by randomly adding
activated monomers to an oligomer as a Poisson process, taking an
experiment on clay surfaces \cite{Kawamura1994} as a model case.
Non-RNA nucleic acid analogues may have carried genetic information
before the RNA world emerged \cite{Ruiz-Mirazo2014}, but the
formulations below can also be applied to such cases.  Let $x_l$ be
the abundance of $l$-nt long oligomers. After the injection of
activated monomers at the time of initialization ($t=0$), evolution of
$x_l$ is described by the following differential equations:
\begin{eqnarray}
  \dot{x}_{l+1} = \kappa \, x_l - \kappa \, x_{l+1} \ ,
  \label{eq:x_l_ev}
\end{eqnarray}
where the dot denotes a time derivative. Here we assume that the
coefficient $\kappa$ (probability of a reaction with a monomer per
unit time) does not depend on the oligomer length, which is
approximately consistent with the trend found in the experiment
\cite{Kawamura1994}. We consider initial conditions of $x_l = 0$ for
$l \ge 2$, and $x_1$ can be approximated to be constant in the early
phase. The second term on the right hand side can be neglected when
$x_{l+1} \ll x_l$.  Solving the equations iteratively under these
conditions, the abundance $x_l$ at a time $t$ is obtained as
\begin{eqnarray}
x_l = \frac{p_r^{l-1}}{(l-1)!} \, x_1 \ ,
\label{eq:poisson}  
\end{eqnarray}
where $p_r \equiv \kappa \, t$ is the reaction probability with a
monomer up to the time $t$.  A similar result is obtained by
considering the Poisson distribution with an expectation value of
$p_r$; the only difference is a factor of $\exp(-p_r)$ that is not
important at $p_r \lesssim 1$.  We should consider only the regime of
$p_r \lesssim 1$, because by the time $t \sim \kappa^{-1}$, a significant
fraction of activated monomers are lost by the reactions, and hence
the approximation of constant $x_1$ is no longer valid and efficient
polymerization is not expected beyond this point.  If activated
monomers are lost earlier by some other processes (e.g. hydrolysis),
$p_r$ would be smaller than unity.

In RNA oligomerization on clay surfaces, the coefficient $\kappa$
should be proportional to the concentration of activated monomers
adsorbed on the clay surface.  This clay-phase monomer concentration
increases with that in aqueous phase, but according to the Langmuir
adsorption isotherm, it saturates when the adsorbed monomer abundance
reaches that of the exchangeable cations on clay surface. In the
experiment \cite{Kawamura1994}, montmorillonite has 0.8 mmol
exchangeable cations per gram, and it starts to saturate at an aqueous
monomer concentration of $\sim 0.01$ M (= mol/L). At the saturated
clay-phase monomer concentration, the reaction rate is $\kappa \sim$ 1
h$^{-1}$, and thus $p_r \sim 1$ is reached within a few hours, which
is much shorter than the hydrolysis time scale of activated
monomers. Aqueous monomer concentration needs to be higher than a
certain level to keep $\kappa$ large enough for $p_r \sim 1$, and this
may be achieved at some points during a cycle, for example, by
variable amount of water expected in dry-wet cycles around warm little
ponds \cite{Pearce2017}.

\section*{Probability of an active RNA formation in the universe}

Let $l_{\min}$ be the minimum length of an RNA that needs to be
abiotically formed for an emergence of life, and suppose that a
$l_{\min}$-nt long, randomly polymerized RNA molecule acquires the
necessary activity with a probability $P_{ac}$ by a correct
informational sequence of nucleotides. Once such an active polymer is
produced, it proceeds to the stage of Darwinian evolution with a
probability $P_{ev}$, thus completing an abiogenesis process.  Then we
can calculate the number of abiogenesis events in a region of the
universe containing $N_*$ stars as
\begin{eqnarray}
  N_{\rm life} = N_* \, f_{pl} \, t_d \, r_p \, P_{ac} \, P_{ev} \ ,
  \label{eq:N_life}
\end{eqnarray}
where $f_{pl}$ is the number of habitable planets per star, $t_d$ the
time during which abiotic RNA polymerization cycles continue, and
$r_p$ the production rate of $l_{\min}$-nt long RNA polymers on a
planet. The production rate by
the Poissonian process can be expressed using eq. \ref{eq:poisson}
in the previous section as
\begin{eqnarray}
  r_p = N_m \, \, \frac{p_r^{l_{\min}}}{l_{\min}!} \, \, t_c^{-1} \ ,
  \label{eq:r_p}
\end{eqnarray}
where $N_m$ is the number of activated monomers participating in a
cycle of polymerization on a planet, $t_c$ the repeating time interval
of polymerization cycles, and an approximation of $l_{\min} \sim
l_{\min} - 1$ is used for simplicity. The baseline value
of $p_r$ is set to unity in the following analysis.

The probability $P_{ac}$ can be expressed as
\begin{eqnarray}
  P_{ac} = \frac{N_{ac}}{N_{nb}^{l_{\min}}} \ ,
\end{eqnarray}
where $N_{nb}$ is the number of nucleobase types participating in
polymerization, and $N_{ac}$ is the number of active sequences among
all the possible sequences of a $l_{\min}$-nt long RNA polymer.  We
adopt $N_{nb} = 4$ as the baseline from RNA/DNA of life as we know it,
but probably this is an underestimate for abiotic polymerization,
because regioselectivity, homochiral selectivity, or any other
reacting molecules that stop further polymerization would effectively
increase $N_{nb}$.  The parameter $N_{ac}$ is highly uncertain. Here
we convert this parameter into $\Delta l$ (or $l_{\rm eff} \equiv
l_{\min} - \Delta l$) defined by the relation $N_{ac} \equiv N_{nb}^{\Delta
  l}$, so that
\begin{eqnarray}
  P_{ac} = \frac{1}{N_{nb}^{l_{\min} - \Delta l}}
  = \frac{1}{N_{nb}^{l_{\rm eff}}} \ .
\end{eqnarray}
Considering an example with $l_{\min} = 40$, there are $4^{40} \sim
10^{24}$ possible sequences of 40-mers, and perhaps $N_{ac} = 10^4$
sequences out of them may have a replicase activity
\cite{Robertson2012}, in this case $\Delta l = 6.6$.  Here we take
$\Delta l = 0$ as the baseline value, which is valid when
$\Delta l \ll l_{\min}$.

Requiring $N_{\rm life} = 1$ and taking a logarithm of
eq. \ref{eq:N_life}, we find the number of stars necessary to expect
at least one abiogenesis event in their planetary systems, as
\begin{eqnarray}
  2.3 \, \lg N_* &=& \ln (l_{\min}!) - l_{\min} \ln p_r
     + \ (l_{\min}-\Delta l) \, \ln N_{nb} - \ln C \ ,
     \label{eq:N_star_RNA}
\end{eqnarray}
where
\begin{eqnarray}
  C \equiv f_{pl} \, N_m \, t_d \, t_c^{-1} \, P_{ev}
\end{eqnarray}
and lg and ln are the common and natural logarithms, respectively. We
need to determine the five parameters included in $C$.  Obviously
there are huge uncertainties, probably more than 10 orders of magnitude
in total.  However, these parameters appear only logarithmically, and
we will find that these uncertainties hardly affect the main
conclusions derived in this work.

We use $f_{pl} = 0.1$ as the baseline for the planet parameter
\cite{Lissauer2014}, which is the least uncertain among these, owing
to the rapid development of exoplanet studies in recent years.  The
baseline for $t_d$ is set to 0.5 Gyr as a plausible time scale from
the birth of Earth to the abiogenesis \cite{Pearce2018}, and that for
$t_c$ is set to 1 yr supposing a seasonal cycle (e.g. ref.
\cite{Pearce2017}), though 1 day may also be reasonable for a
day-night cycle.  The parameter $P_{ev}$ is highly uncertain, but
$P_{ev} = 1$ is set as the baseline, which is optimistic but may not
be unreasonable because a long RNA polymer assembled by the Poisson
process would be rare and there would be no competitor or predator
around it.  Any other essential factors involved in the origin of
life, e.g., encapsulation by membrane vesicle formation, may
significantly reduce this parameter. It has been known that both RNA
polymerization and vesicle assembly are accelerated on clay surfaces
\cite{Ferris2006,Ruiz-Mirazo2014,Kitadai2018}.

The amount of monomers, $N_m$, is probably the most uncertain
parameter among the five in $C$.  An upper limit may be estimated by
the number of nucleotides in the present life on Earth, $N_m = 7
\times 10^{37}$ ($3.7 \times 10^{16}$ g in mass), which is estimated
by the total biomass of 550 Gt-C ($3.7 \times 10^{12}$ wet$\cdot$t)
\cite{Bar-On2018} assuming that nucleic acids constitute 1\% of the
wet biomass.  A rough amount of nucleobases delivered from space by
meteorites can be estimated as follows (see ref.\cite{Pearce2017} for a
more detailed modeling).  The mass delivery rate of meteoroids from
4.5 to 4.0 Ga is $10^{20-25}$ kg/Gyr, and 0.1\% of this mass belongs
to meteoroids of a diameter 40--80 m, which efficiently deliver
nucleobases avoiding melting or vaporization.  Carbonaceous meteorites
contain nucleobases with a mass fraction of $10^{-7}$, and they are
deposited into warm little ponds which cover a fraction of $10^{-6}$
on the Earth surface. These nucleobases survive for 1 yr, i.e., a
seasonal cycle before they are destroyed by UV radiation or
seepage. Then we expect $10^{20-25}$ nucleobases (0.01--$10^3$ g in
mass) in the ponds. Instead, nucleobases may also be produced on
Earth, and it would not be unreasonable to assume a similar
nucleobase/carbon mass ratio to that found in carbonaceous meteorites
($10^{-5}$). Assuming a carbon abundance similar to the present
seawater, we expect $10^{27}$ nucleobases in the ponds assuming their
depth to be 1 m.  We then use $N_m = 10^{25}$ as the baseline, though
it could be wrong by many orders of magnitude, depending on various
scenarios of nucleotide formation and their activation under prebiotic
conditions.  Using the baseline parameter values thus determined, we
find $\ln C = 75.3$.

\section*{The minimum RNA length versus the universe size}

\begin{figure*}[t]
\centering
\includegraphics[width=15cm,angle=0]{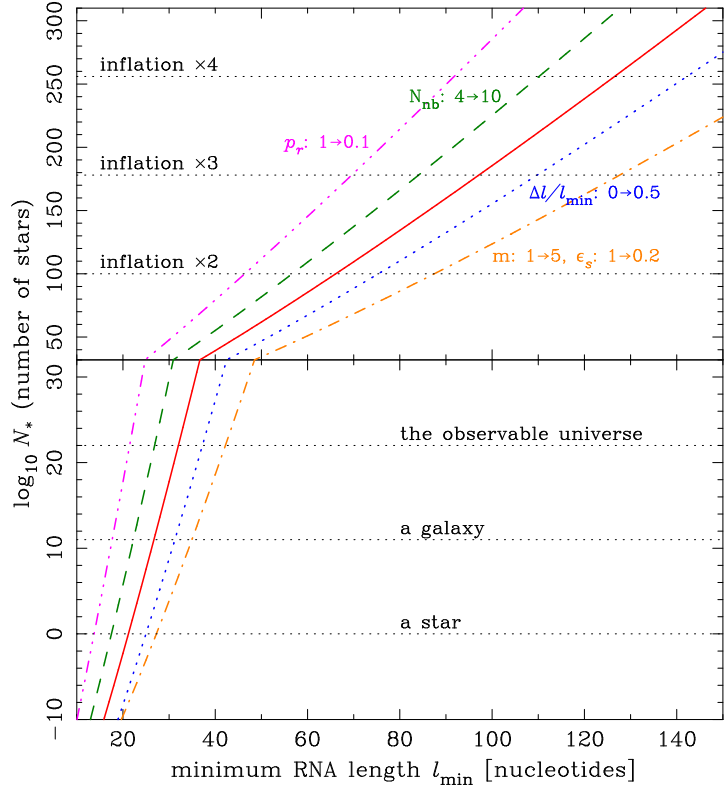}
\caption{ Logarithm of the number of stars necessary to expect at
  least one abiogenesis event ($\lg N_*$) versus the minimum RNA
  length required to show a biological activity leading to abiogenesis
  ($l_{\min}$). The difference between the top and bottom panels is
  just the scale of the vertical axis.  Some important values of $\lg
  N_*$ are indicated by horizontal dotted lines; ``inflation $\times
  2$'' means the universe size when the inflation lasted twice as long
  as that required to make the observable universe homogeneous. The
  red solid curve is the relation using the baseline model parameter
  values, and other curves are when some of the model parameters are
  changed from the baseline values, as indicated in the figure.}
\label{fig:RNA_star}
\end{figure*}

Fig. \ref{fig:RNA_star} shows $\lg N_*$ versus $l_{\min}$ for $N_{\rm
  life}= 1$ calculated by eq. \ref{eq:N_star_RNA}.  When the baseline
parameter values are used, the minimum RNA length must be $l_{\min} =
21, 27$ and $32$ to expect one abiogenesis event for a survey of a
single star ($\lg N_* = 0$), a galaxy ($\lg N_* = 11$), and the
observable universe ($\lg N_* = 22$), respectively.  These $l_{\min}$
values are not sufficiently large compared with that ($\sim$40--100)
required to expect an RNA replicase activity from a biological
viewpoint, implying that abiogenesis is not easy even if we consider
the entire volume of the observable universe.  For $l_{\min} = 40$ we
find $\lg N_* = 39$.  If we try to reduce this to $\lg N_* = 22$ or 0
for the same $l_{\min}$ by the uncertainty in $C$, this parameter
needs to be increased by a factor of $10^{17}$ or $10^{39}$,
respectively.

However, if we request one abiogenesis event somewhere in the whole
physical volume created by an inflation, the chance of abiogenesis
greatly increases. In a volume created by a twice, three, and four
times as long inflation as that required to create the observable
universe ($\lg N_* = 100, 178$, and 256), $l_{\min}$ becomes 66, 97,
and 127, respectively.  These $l_{\min}$ lengths now allow us to
expect a self-replicating activity of an RNA molecule.  If an
identical pair of RNA strands is required for abiogenesis, $l_{\min}$
should be effectively twice as large as each of the identical strands.
Then an inflationary universe can produce a pair with a length of
$\sim$33--64 nt for each, and we can still expect a replicase
activity. It is also possible that the inflation duration is even
longer than the examples considered here.

Using the Stirling's approximation, 
eq. \ref{eq:N_star_RNA} can be written as
\begin{eqnarray}
2.3 \, \lg N_* \sim l_{\min} \, ( \, \ln l_{\min} + \ln N_{nb} -1) - \ln C 
\end{eqnarray}
when $p_r = 1$ and $\Delta l = 0$. In the large limit
of $l_{\min}$, $\ln N_{nb} - 1$ can be neglected, and a
useful approximated formula is:
\begin{eqnarray}
\lg N_* \sim  l_{\min} \, \lg l_{\min} - \lg C \ .
\end{eqnarray}
It should be noted that $\lg N_*$ changes only by 10 when a factor
included in $C$ is changed by 10 orders of magnitude; $\lg N_*$
changes from 167 to 177 for $l_{\min} = 100$ for example,
which hardly affects the main conclusion of this work. 

Fig. \ref{fig:RNA_star} also shows the $\lg N_*$-$l_{\min}$ relation
when some model parameters are changed. The main results described
above are not seriously changed when we change $N_{nb} = 4 \rightarrow
10$ or $\Delta l / l_{\min} = 0 \rightarrow 0.5$. If we reduce $p_r$ from 1
to 0.1, $l_{\min}$ is reduced from 66 to 46 for $\lg N_* = 100$ (a
twice as long inflation). A sufficient number of abiogenesis events may
not be expected when $p_r \ll 1$, even in the total volume of an
inflationary universe.

A possibly important process is polymerization over multiple cycles.
In polymerization on clay surfaces, inactive monomers and oligomers
left from the previous cycle must be released from a clay surface for
the next cycle to work, but a fraction of long oligomers may remain on
the surface. Adding newly activated monomers to such oligomers over
many cycles may be an efficient way to assemble a long polymer. Such a
polymerization process may be limited by a time scale of RNA oligomer
destruction, e.g., by hydrolysis or UV radiation during the dry phase.
As a toy model to consider this, suppose that a fraction $\epsilon_s$
of oligomers survive to the next cycle. If polymerization of an oligomer
continues over $m$ cycles, the most efficient path to form a
$l_{\min}$-nt polymer would be to repeat $m$ times the process of
adding $l_{\min}/m$ monomers. Then the polymer production rate $r_p$
of eq. \ref{eq:r_p} should be replaced by
\begin{eqnarray}
  r_p = N_m \left[ \frac{ p_r^{l_{\min}/m}  }{(l_{\min}/m)!} \right]^m
  \, \epsilon_s^m (m \, t_c)^{-1} \ .
\end{eqnarray}
The result for $m = 5$ and $\epsilon_s = 0.2$ is shown in
Fig. \ref{fig:RNA_star} as an example, using the Gamma function for
the factorial when $l_{\min}/m$ is not an integer.  In this case
$l_{\min}$ becomes 42 for $\lg N_* = 22$, implying a possibility that
abiogenesis has occurred more than once inside the observable
universe.  Though $m$ and $\epsilon_s$ are highly uncertain, this
possibility should not be overlooked.

\section*{Conclusions}

It has been shown that the first RNA polymer with a replicase activity
can be abiotically assembled by the most conservative polymerization
process, i.e., random Poissonian adding of monomers, if we require
that it occurs more than once somewhere in the physical volume of a
universe created by an inflation, rather than inside the observable
universe for us. This gives a simple solution to the abiotic
polymerization problem to initiate the RNA world.  Equation
\ref{eq:N_star_RNA} relates two quantities on vastly difference
scales: $\lg N_*$ on an astronomical scale and $l_{\min}$ on a
biologically microscopic scale, and uncertainties of other parameters
are not important because most of them appear logarithmically.  This
reminds us of an ouroboros.

The result of this work may also give an explanation for the
homochirality of life. Even if activated monomers supplied to the
polymerization cycle are a racemic mixture, life emerging from them
would be homochiral, if homochirality is a necessary requirement for
an RNA polymer to show biological activities. Simply it needs more
time or volume for a homochiral polymer to be assembled by random
polymerization, with $N_{nb}$ twice as large as when ignoring
chirality. As shown in Fig.  \ref{fig:RNA_star}, change of $N_{nb}$ by
a factor of two does not seriously affect the expected number of
abiogenesis events in an inflationary universe.

On the other hand, the expected number of abiogenesis events is much
smaller than unity when we observe a star, a galaxy, or even the whole
observable universe. This gives an explanation to the Fermi's paradox.
The observable universe is just a tiny part, whose volume is likely
smaller than $1/10^{78}$ of the whole universe created by an
inflation, and there is no strong reason to expect more than one
abiogenesis event in such a small region.  Even if Earth is the only
planet that harbors life inside the observable universe, it does not
contradict the Copernican principle, because life would have emerged
on countless planets in the whole inflationary universe in which we
exist.

In the picture presented here, the probability of finding
biosignatures from planets or satellites in the Solar System or from
exoplanets is negligibly small, unless we consider interplanetary or
interstellar traveling of microorganisms
\cite{Nicholson2009,Wesson2010}. It should be noted, however, that the
case of a high abiogenesis rate ($N_{\rm life} \gtrsim 1$ for $N_* =
1$) cannot be excluded by this work, because we assumed that abiotic
RNA polymerization occurs only by the random Poisson process of adding
monomers.  Potential roles of much more efficient processes on the
origin of life, such as non-linear auto- or cross-catalytic reactions,
have been studied theoretically \cite{Coveney2012}, though it is
highly uncertain whether such processes really worked in realistic
prebiotic conditions. If organisms having a different origin from
those on Earth are found in future, it would suggest that such a
mechanism is working at abiogenesis to reduce $l_{\min}$.  Although
this possibility should not be excluded, what is shown by this work is
that such a hypothetical process is not necessary if we request
abiogenesis events to occur somewhere in an inflationary universe.

It is also worth pointing out that, in the $\lg N_*$-$l_{\min}$
relation for $N_{\rm life} = 1$, $\lg N_*$ rapidly increases from 0 (a
star) to 22 (the observable universe) in a short range of $l_{\min} =
$ 21--32. Even if a non-linear process is working at some stages, the
initial polymerization is likely statistical and random as considered
here. Then it would be an extreme fine tuning if a biological
parameter $l_{\min}$ is just close to the value corresponding to
$N_{\rm life} \sim 1$ for a star ($N_* = 1$). Rather, $N_{\rm life}
\gg 1$ or $N_{\rm life} \ll 1$ is much more likely when we observe
just one planetary system. As we have argued, the case of $N_{\rm
  life} \ll 1$ is not in contradiction with observations, but the
opposite case may be in tension with the lack of evidence for multiple
abiogenesis events in the history of Earth or in laboratories.

A fundamental assumption in this work is that an abiotically assembled
RNA polymer acquires a self-replicating ability if it is sufficiently
long and has a correct nucleotide sequence. This may be rather trivial
under the physical laws ruling this universe, because we know that
ribozymes are actually working in life and can also be produced by
{\it in vitro} experiments.  This work considered only a single
homogeneous region in the universe created by an inflation event,
obeying the same physical laws that we observe. However, the multiverse
hypothesis \cite{Carr2009} implies existence of other universes
created by different inflation events, in which physical laws may be
different from ours.  A theoretically intriguing question is whether a
chemical RNA-like long polymer is easily formed to contain information
and show biological activities eventually leading to higher organisms,
when physical laws are arbitrarily made, e.g., by random choices of
fundamental physical constants.  Perhaps this may be the ultimate
mystery regarding the origin of life, which is, of course, far beyond
the scope of this work.

\bibliography{totani}

\section*{Acknowledgements (not compulsory)}
The author was supported by the JSPS/MEXT KAKENHI Grant Numbers
18K03692 and 17H06362.

\section*{Author contributions statement}

T.T. performed all work for this paper.

\section*{Additional information}


\textbf{Competing interests}

The author declares no competing interests.

\end{document}